\documentclass[11pt, notitlepage, letterpaper]{article}
\pdfoutput = 1
\usepackage[utf8]{inputenc}
\usepackage{booktabs}
\usepackage[ruled]{algorithm2e}
\usepackage{amsfonts, amsmath, amsthm, mathtools, xcolor, hyperref, graphicx, enumerate, dsfont, csquotes, paralist, bbm, url, thmtools, amssymb, comment, cleveref, algorithmicx}
\usepackage[english]{babel}
\usepackage[nice]{nicefrac}
\usepackage[margin=1in]{geometry}

\usepackage[sorting=ynt, url = false, doi = false, isbn =false, backend = biber]{biblatex}
\SetAlFnt{\small}
\SetAlCapFnt{\small}
\SetAlCapNameFnt{\small}
\SetAlCapHSkip{0pt}
\IncMargin{-\parindent}
\allowdisplaybreaks
\newtheorem{theorem}{Theorem}

\newtheorem{proposition}{Proposition}
\newtheorem{lemma}{Lemma}
\newtheorem{claim}{Claim}

\newtheorem*{theorem*}{Theorem}

\SetAlCapNameFnt{\bfseries} 
\SetAlCapFnt{\bfseries}

\newcommand{\EX}{\mathbb{E}}

\title{Static Pricing for Single Sample Multi-unit Prophet Inequalities\thanks{A preliminary version of the first result in this paper first appeared in the report `Estimated Prophet: The Prophet Inequality in Fair, Sample, and Multi-Selection Settings', by Q McIntyre, E Steinberg, P Westbrook, and E Zhang available here: \url{https://surim.stanford.edu/sites/g/files/sbiybj26191/files/media/file/pranav_group_1.pdf}}}

\author{Pranav Nuti
\thanks{Booth School of Business, University of Chicago, \texttt{pranavn@uchicago.edu}}
\and
Peter Westbrook
\thanks{Citadel Securities, \texttt{petewest@stanford.edu}}
}
\date{}
\begin{document}

\maketitle

\begin{abstract}
In this paper, we study $k$-unit single sample prophet inequalities. A seller has $k$ identical, indivisible items to sell. A sequence of buyers arrive one-by-one, with each buyer's private value for the item, $X_i$, revealed to the seller when they arrive. While the seller is unaware of the distribution from which $X_i$ is drawn, they have access to a single sample, $Y_i$ drawn from the same distribution as $X_i$. What strategies can the seller adopt for selling items so as to maximize social welfare?

Previous work has shown that when $k=1$, if the seller sets a price equal to the maximum of the samples, they achieve a competitive ratio of $\frac{1}{2}$ of the social welfare, and recently Pashkovich and Sayutina established an analogous result for $k=2$. In this paper, we prove that for $k\geq 3$, setting a static price equal to the $k^{\text{th}}$ largest sample also obtains a competitive ratio of $\frac{1}{2}$, resolving a conjecture Pashkovich and Sayutina pose.

We also consider the case where $k$ is large. We show that setting the $(k-\sqrt{2k\log k})^{\text{th}}$ largest sample as the price obtains a competitive ratio of $1-\sqrt{\frac{2\log k}{k}}-o\left(\sqrt{\frac{\log k}{k}}\right)$, and that this is optimal for static pricing with access to a single sample. This should be compared against $1-\sqrt{\frac{\log k}{k}}-o\left(\sqrt{\frac{\log k}{k}}\right)$, the ratio obtainable using static pricing with knowledge of the distributions of the values.


\end{abstract}

\thispagestyle{empty}
\newpage
\setcounter{page}{1}

\section{Introduction and motivation}

Consider a seller who wishes to sell a single, indivisible item to a sequence of buyers who arrive one-by-one. Upon the arrival of a buyer, the buyer's private value is revealed to the seller, who must then immediately and irrevocably decide whether to allocate the item to the current buyer.  How should the seller decide who to allocate the item to so as to maximize social welfare?

If the seller has access to independent distributions from which the buyers' valuations are drawn, this is a problem investigated by the classic \textit{prophet inequality}. The name comes from imagining the seller competing with, and trying to obtain a fraction of, the value obtained by a \textit{prophet} who knows all of the buyers' private valuations (and does not need to observe these valuations sequentially). The fraction obtainable is called the \textit{competitive ratio}.

Prophet inequalities were introduced by Krengel and Sucheston \cite{KS77} as an optimal stopping problem, and they have received significant attention from probabilists. For a survey of a lot of the early work related to prophet inequalities, see \cite{HK92}. More recently, since the work of \cite{HKS07} and \cite{CHMS10} noticing the connection of the classic prophet inequality with allocation schemes, prophet inequalities have also been of interest to economists and computer scientists.

One significant achievement of the early work on the prophet inequality was the demonstration that there exists a simple take-it-or-leave-it price that the seller can set to obtain a competitive ratio of $\frac{1}{2}$ of the prophet, and that this is the best possible ratio obtainable by any allocation scheme \cite{samuel1984comparison}.

It is natural to ask about what happens when the seller has $k$ identical items they wish to sell (and each buyer is interested in buying one). \cite{HKS07} investigated static pricing strategies for this problem, and demonstrated the existence of a price which obtains a competitive ratio of $1 - O\left(\sqrt{\frac{\log k}{k}}\right)$. More recently \cite{CDL23} found the optimal static pricing strategy for the problem, and \cite{JMZtight} demonstrated that this strategy was indeed optimal.

There has also been significant interest in investigating more general strategies for allocation, with a particularly important line of work related to \textit{online contention resolution schemes} (OCRSs). \cite{alaei2014bayesian} demonstrated the existence of strategy for the seller which guarantees a competitive ratio of $1 - \frac{1}{\sqrt{k+3}}$, and this factor was further improved by \cite{JMZocrs}, who showed that the OCRS in \cite{alaei2014bayesian} was optimal. While OCRSs perform better than what is achievable with a static pricing strategy, from the perspective of applications, this comes at a significant cost---a take-it-or-leave-it price is simple, anonymous, and order-oblivious (see the survey \cite{luciersurvey} for a discussion of these economic advantages).

All of this work assumes that the seller has access to the distributions from which the buyers' private valuations are drawn. But this is perhaps unrealistic. A more limited assumption is that the seller has access to just a single sample from each valuation distribution (perhaps from having sold to the same buyers before). Under this seemingly extremely restrictive assumption, \cite{AKW14} showed using the \textit{rehearsal algorithm} that the seller can obtain a competitive ratio of $1 - O\left(\frac{1}{\sqrt{k}}\right)$, nearly matching what is achievable with full knowledge of the distribution.

Unfortunately, the rehearsal algorithm is not a static pricing strategy, and as such, enjoys none of the advantages that come from a simple take-it-or-leave-it price. We are therefore interested in the following question:

\begin{displayquote}
   \emph{In case a seller wishing to sell $k$ items has access to just a single sample from each distribution from which the buyers' valuations are drawn, what competitive ratio can be achieved by a static pricing strategy?}
\end{displayquote}

This question has been previously answered elegantly in the case $k = 1$ in \cite{RWW20}. They show that setting a price equal to the largest observed sample obtains a competitive ratio of $\frac{1}{2}$, matching what is possible with full knowledge of the distributions of the valuations. More recently, \cite{PS23} investigated the question for $k = 2$, and showed that setting a price equal to the second largest observed sample obtains a competitive ratio of $\frac{1}{2}$. This paper studies the case $k \geq 3$.

\subsection{Our results}

To explain our results, we begin by quickly establishing some notation. In  this paper, we work with a sequence of independent continuous random variables $X_1, X_2, \ldots, X_n$ (with $X_i \sim \mathcal{D}_i$) representing the $n$ buyers' valuations. Furthermore, we assume that for each $X_i$, we (taking on the role of the seller) have access to an independent sample $Y_i \sim \mathcal{D}_i$. The $Y_i$ are available to us all at once, before we have to make any decisions to sell. It will be convenient to write $X^j$ for the $j^{\text{th}}$ largest number among the $X_i$.\footnote{This notation is non-standard. The usual notation $X_{i:n} = X_{(i)}$ for the $i^{\text{ th }}$ smallest order statistic is more inconvenient when, as in this paper, we primarily focus on the top few order statistics, and $n$, the number of random variables, is mostly irrelevant.} We also let $X^{j}=0$ for $j > n$ and $X^0 = \infty$.\footnote{For the purpose of analysis, ties between $0$s are broken randomly.}  Define $Y^j$ analogously.

We are interested in algorithms that observe the $X_i$ in an online fashion, and make $k \leq n$ selections from the $X_i$, aiming to maximize their sum. We study the following algorithm which sets the  $r^{\text{th}}$ largest sample as price:
\begin{algorithm}[ht]
\caption{Static pricing for single sample multi-unit prophet inequalities}
Input: $Y_i\sim  \mathcal{D}_i$ \\
$j \leftarrow 0$\\
ALG $\leftarrow 0$\\
\For{i=1, \ldots, n}{
    \If{$X_i>Y^r$ and $j<k$}{
    ALG $\leftarrow$ ALG $+ X_i$, $\text{ i.e., sell to } t \text{ and add } X_i \text{ to total social welfare}$\\
    $j \leftarrow j+1$
    }
}
\label{offlinek-select}
\end{algorithm}
\begin{theorem}
    For any $k$, Algorithm \ref{offlinek-select} with $r = k$ is $\frac{1}{2}$-competitive, that is, 
    $$ \mathbb{E}[ALG] \geq \frac{1}{2} \mathbb{E}\left[\sum_{i=1}^k X^i\right].$$ 
    \label{k-select}
\end{theorem}
 Pashkovich and Sayutina proved this result for $k=2$ \cite{PS23}, and conjectured that it was true for higher values of $k$.\footnote{Without assuming that the random variables are continuous, we can also recover the result using randomized tie-breaking (see \cite{RWW20}) whenever two variables have the same realization. However, if we ask about completely deterministic algorithms, it is not clear whether there is an algorithm with a competitive ratio of $\frac{1}{2}$, even for $k=2$.} Theorem \ref{k-select} resolves their conjecture, and in fact, we provide a simpler proof, even for $k = 2$. Note that setting the $k^{\text{th}}$ largest sample as price is not necessarily the best choice, but the simplicity of this algorithm, and the concrete bound we can obtain for small values of $k$ makes it interesting to analyze. Note that no such concrete bound was known earlier, even for $k = 3$.\footnote{ We note also that it is easy to see that the factor of $\frac{1}{2}$ in \cref{k-select} is tight by considering the case where $X_i\sim \text{Unif}(1-\delta,1)$ when $ 1 \leq i\leq n-1$, while  $X_n \sim \text{Unif}(1-\delta,1)$ with probability $\frac{n-1}{n}$ and $n^2$ with probability $\frac{1}{n}$ with $n$ large.}

We also investigate the case of large values of $k$, and prove:
 \begin{theorem}
Algorithm \ref{offlinek-select} with $r \leq k$ has a competitive ratio of at least: 
\[\frac{r\Pr(Z \leq k-1) - r\frac{\binom{2r-1}{r}}{2^{2r-1}}}{k},\]
where $Z\sim \text{Binom}(r+k-1, \frac{1}{2})$. In particular, if we set $r = k -\sqrt{2k\log k}$, we obtain a competitive ratio of at least $1 - \sqrt{\frac{2\log k}{k}} - O\left(\frac{1}{\sqrt{k}}\right)$, that is, 
    $$ \mathbb{E}[ALG] \geq \left(1 - \sqrt{\frac{2\log k}{k}} - o\left(\sqrt{\frac{\log k}{k}}\right)\right) \cdot \mathbb{E}\left[\sum_{i=1}^k X^i\right].$$
    \label{k-select-large}
\end{theorem}

(Throughout, the logarithms we use have base $e$.)\footnote{For a fixed $r$ and  $k$, our result is not optimal. In fact, it is also possible to obtain a slightly better competitive ratio of $\frac{r\Pr(Z \leq k-1)}{k} \left(1 - \frac{\binom{2r-1}{r}}{2^{2r-1}}\right)$, but we omit the proof for the sake of simplicity, since this improved factor also seems to be non-optimal.} While algorithms which obtain a better competitive ratio were known earlier, either they were not static pricing strategies, or they had access to the entire distribution from which the valuations are drawn.\footnote{A static pricing strategy similar to our own is also considered in \cite{HKS07}, where knowledge of the entire distribution (rather than a single sample) is used to set a deterministic price $T$ which ensures that $\sum_{i = 1}^n\Pr(X_i \geq T) = k -\sqrt{2k\log k}$. Their analysis is not tight.} The performance of our algorithm should be compared against the performance of the optimal algorithm with access to the entire distribution (which obtains a competitive ratio of $1 - \sqrt{\frac{\log k}{k}} - o\left(\sqrt{\frac{\log k}{k}}\right)$, as we show in \cref{sec:old-result}).

Finally, we can also demonstrate that \cref{k-select-large} is tight:
\begin{theorem}
    The lower bound of Theorem \ref{k-select-large} is the best achievable with any (potentially randomized) static pricing strategy with access to a single sample from each distribution (up to the factor of $o\left(\sqrt{\frac{\log k}{k}}
    \right)$).  
    \label{tightness}
\end{theorem}
Note that we do not require in \cref{tightness} that the algorithm is necessarily one that uses one of the samples as price. In other words, it is not useful to look at the exact values of the samples, and an algorithm that uses just comparisons obtains the optimal competitive ratio. Some recent work has explored the optimality of ordinal algorithms for ordinal problems \cite{GST23}. These results do not seem to apply directly in our setting, and our proof is significantly simpler.

\subsection{Our techniques}

Prior to our work, it was known \cite{CDL23} that by setting a deterministic price of $p$, it is possible to obtain a competitive ratio of:

\[\min\left\{\Pr\left(X^k < p\right), 
  \frac{\sum_{i=1}^k \Pr(X^i \ge p)}{k}\right\}.\]

The two quantities in the minimum correspond to two hard instances. $\Pr\left(X^k < p\right)$ is what we would obtain if all the random variables were roughly the same size, except for the last one, which has a small chance of being really, extremely large. Our competitive ratio is then dominated by the probability with which we manage to pick the last random variable. The danger in this instance is that we set too low a price, and select $k$ of the first random variables, not saving up enough of a probability for picking the last variable.

$\frac{\sum_{i=1}^k \Pr(X^i \ge p)}{k}$ is what we would obtain if all the random variables have roughly the same size. Our competitive ratio is basically then the same as the number of picks we make. The danger in this instance is that we set too high a price, and don't select enough of the random variables.

In our work, we are interested in setting a price equal to $Y^r$, the  $r^{\text{th}}$ largest observed sample. If we were to just utilize the result of \cite{CDL23}, the resulting competitive ratio would be:

\[\EX_{p \sim Y^r}\left[\min\left\{\Pr\left(X^k < p\right), 
  \frac{\sum_{i=1}^k \Pr(X^i \ge p)}{k}\right\}\right].\]

Unfortunately, the two terms in the minimum are negatively correlated with each other, which has a tendency to push the expected value of their minimum down, and makes it hard to compute. Fortunately, this is not a serious impediment to the analysis when $r$ and $k$ are large, and $r$ is somewhat separated from $k$--all we need to do is to lower bound the minimum by the sum of the two quantities minus one. Taken together with estimates on $\Pr(X^i > Y^r)$, this completes the proof of \cref{k-select-large}.

The proof of \cref{tightness} depends on the same hard instances we've just discussed, but requires a little care because we are allowing for a general class of cardinal algorithms that can use the samples in potentially arbitrary, randomized ways to set thresholds.

However, the proof of \cref{k-select} requires completely different ideas, because in this case, we are trying to establish the optimal competitive ratio when $r = k$, and there is a loss in the analysis of \cref{k-select-large} when we attempt to estimate the correlation between the two terms in the minimum.

In order to demonstrate that our algorithm for $r=k$ is $\frac{1}{2}$-competitive, we start by adopting a stochastic dominance viewpoint commonly used for establishing various prophet inequalities \cite{AKW14, Googol}, and concentrate on showing that, for every $x$, the expected number of picks our algorithm makes above $x$ is at least $\frac{1}{2}$ times the expected number of picks the prophet makes above $x$.
 
Next, we use a standard method to analyze single sample prophet inequalities and related problems \cite{AKW14, RWW20, Googol, PS23} that exploits the symmetry between the $X_i$ and $Y_i$. We imagine that for each $i$, a pair of realizations are drawn from $\mathcal{D}_i$, and one of them is randomly assigned (with probability $\frac{1}{2})$ to $X_i$ and the other to $Y_i$. We then condition on the pair of realizations and assume that they are fixed throughout the argument, and the only randomness that remains involves which one of the two realizations is assigned to $X_i$ and which to $Y_i$. We collect realizations from all the $\mathcal{D}_i$ into a ranked list: $\{a_1 > a_2> \ldots > a_{2n}\}$, with each $a_l$ being assigned to $X$ or $Y$ randomly. 

If $a_l$ is assigned to $X$, then some paired $a_m$ is definitively assigned to $Y$, so there is a correlation in how the $a_l$ are assigned. Previous work has had to carefully account for which $a_l$ and $a_m$ are paired in this way \cite{PS23}. We are able to almost completely avoid this and significantly simplify previous work by treating the problem more abstractly. We do this by proving \cref{1/2 Lemma} and \cref{lemma:inbetweenrank}, key de-correlation lemmas that capture all the relevant information about the correlation and the pairing.\footnote{Another de-correlation lemma that shares some similarities in spirit with our own is Lemma 8 in \cite{AKW14}.} 

\subsection{Further related work}

It is possible to make alternative assumptions about the buyers' valuations, and the order in which they arrive.  \cite{correaiid} finds the optimal competitive ratio in case the valuations' are identically distributed. Settings where the buyers arrive in a random rather than in an adversarial order have been studied under the name of \textit{prophet secretary} problems \cite{prophetsecretary}. Static threshold policies for the multi-unit prophet secretary problem have recently been investigated in \cite{arnostima}; they find the optimal static price for the problem.

Prophet inequalities have also now been studied in a variety of contexts where there are more complicated constraints about which values can be accepted (see \cite{kleinberg2019matroid} for work on matroids, \cite{gravin2019prophet} for work related to matchings), or contexts where there are buyers with more complicated preferences (see \cite{FGL15} for a discussion of combinatorial auctions). Some of these settings have also been explored under the assumption of access to just a single sample \cite{singlesamplegreedy}.

For a more extensive survey of related work, see \cite{correasurvey}.

\section{The $k^{\text{th}}$ sample as price}

In this section, we will establish Theorem \ref{k-select} for Algorithm \ref{offlinek-select}. This has already been proven for the case where $k=1$ \cite{RWW20}, and $k = 2$ \cite{PS23}. Here we will prove the result for all $k \geq 1$. 

\subsection{Preliminaries}

In the following, we will refer to realizations of the $X_i$ random variables as $X$-values, or simply $X$'s, and the $Y_i$ random variables as $Y$-values, or simply $Y$'s. We will also say that $X^1, X^2, \ldots, X^k$ are the prophet's picks.

Note that the value the algorithm obtains is clearly more than the sum of the $k$ lowest $X$ values that are above $Y^r$, i.e., if $X^t>Y^r$ and $Y^r>X^{t+1}$, then:

\[
ALG \geq \sum_{i = 1}^{\min\{t, k\}} X^{t - i + 1}.
\]

Hereafter, when we refer to the algorithm's performance, we will implicitly assume that it obtains the right hand side (rather than the left hand side) of the above inequality. When we talk about the algorithm's picks, we will mean the $X^{t-i+1}$s.

Throughout, as we discussed in the introduction, we will condition on $\{a_1 > a_2> \ldots > a_{2n}\}$, the ranked list of the realizations $\{X_1,\ldots, X_n, Y_1, \ldots, Y_n\}$ (and implicitly let $a_m = 0$ for $m > 2n$). We will assume that these values are fixed, but whether each $a_i$ is an $X$ or a $Y$-value is random. Writing $a_i \in X$ to mean that $a_i$ is a $X$-value, $\Pr(a_i \in X) = \Pr(a_i \in Y) = \frac{1}{2}$. Note that $a_i$ being a $X$-value and $a_j$ being a $X$-value are not necessarily independent events. Indeed, if $a_i$ and $a_j$ come from the same distribution, then at most one of them is an $X$-value. Each $a_l$ is paired with some $a_m$ in this way.

In order to prove theorem~\ref{k-select}, we can express the algorithm's expected performance as 
$$\sum_{i=1}^k \EX[A_i],$$
where $A_i$ denotes the algorithm's $i$th largest pick, if any. Thus, the algorithm's expected performance is
$$\sum_{i=1}^k \int_0^{\infty}\Pr(A_i\geq x) dx,$$
and the prophet's becomes 
$$\sum_{i=1}^k \EX[X^i]=\sum_{i=1}^k \int_0^{\infty}\Pr(X^i\geq x)dx.$$ 

Conditioning on and fixing $\{a_1 > a_2> \ldots > a_{2n}\}$, the ranked list of the realizations $\{X_1,\ldots, X_n, Y_1, \ldots, Y_n\}$, in order to establish a competitive ratio of $\frac{1}{2}$, it suffices to show, for all $j\leq 2n$, that 
$$ \sum_{i=1}^k \Pr(A_i\geq a_j)\geq \frac{1}{2} \sum_{i=1}^k \Pr(X^i\geq a_j).$$
The sum on the left is the expected number of the algorithm's picks above $a_j$, so it in fact suffices to show 
\[\EX [\text{Number of ALG picks} \geq a_j]\geq \frac{1}{2} \sum_{i=1}^k \Pr(X^i\geq a_j).\]

But note that $X^i \geq a_j \iff X^i > Y^{j+1-i}$. The above then reduces to

\[\EX [\text{Number of ALG picks } \geq a_j]\geq \frac{1}{2} \sum_{i=1}^k \Pr(X^i > Y^{j+1-i}).\]

Establishing this inequality will be our goal in the next section. The proof of Theorem~\ref{k-select} also depends on two intuitive probabilistic inequalities. Here, we state these lemmas, and we postpone their proofs to the appendix.

\begin{lemma}\label{1/2 Lemma} If $j - k < i \leq j$, then \[\Pr(X^i>Y^{j+1-i}, Y^k>X^{i+k})\geq \frac{1}{2}\Pr(X^i>Y^{j+1-i}).\]
\end{lemma}
This lemma says that if we condition on $X^i \geq a_j$, or in other words, if we condition on which of the $a_l$'s is $X^i$, it is more likely that not that we see $k$ $Y$-values before we see $k$ \textit{additional} $X$-values after $X^i$. Roughly speaking, this is a direct consequence of the fact that each $a_l$ smaller than $X^i$ that is paired with $a_m$ also smaller than $X^i$ is equally likely to be an $X$-value of a $Y$-value.

\begin{lemma}    \label{lemma:inbetweenrank}
If $i \leq j \leq k$, then
\[\Pr(X^i > Y^j > X^k) \geq \Pr(X^i > Y^k > X^k).\]
\end{lemma}

This lemma says that it more likely that $Y^j$ lies in the interval $(X^i, X^k)$ than $Y^k$. In fact, we prove something stronger--the chance that $Y^j$ lies in the interval $(X^i, X^k)$ is actually unimodal.

We are now ready to prove \cref{k-select}.

\subsection{Proof of \cref{k-select}}

\begin{proof}Recall that it suffices to show that for any $j$ (possibly smaller than $k$) that, 
\begin{equation}
  \EX [\text{Number of ALG picks } \geq a_j]\geq \frac{1}{2} \sum_{i=1}^k \Pr(X^i > Y^{j+1-i}).  \label{maininequality}
\end{equation}
The expectation on the left can be expressed as the sum of probabilities from $i=1$ to $n$ that $X^i\geq a_j$ and the algorithm picks $X^i$. It is guaranteed that $X^i\geq a_j$ and the algorithm picks $X^i$ when $X^i>Y^{j+1-i},$ $Y^k>X^{i+k}$ and $X^i > Y^k$, since in this case, $X^i$ lies in the first $k$ $X$-values above $Y^k$. Therefore, 
\begin{align*}
    \EX [\text{Number of ALG picks } \geq a_j]&\geq \sum_{i=1}^n \Pr(X^i>Y^{j+1-i}, Y^k>X^{i+k}, X^i > Y^k)\\
    &= \sum_{i=1}^{j-k} \Pr(X^i>Y^k>X^{i+k})+\sum_{i=j+1-k}^n \Pr(X^i>Y^{j+1-i}, Y^k>X^{i+k}),
\end{align*}

An issue with dealing with the second sum above is that the two events $\{X^i>Y^{j+1-i}, Y^k>X^{i+k}\}$ are correlated. By \cref{1/2 Lemma}, we can remove this correlation, making the task of estimating the expected number of picks of the algorithm significantly simpler. We conclude that:
\begin{equation}
\EX [\text{Number of ALG picks } \geq a_j]\geq \sum_{i=1}^{j-k} \Pr(X^i>Y^k>X^{i+k})+\frac{1}{2}\sum_{i=j+1-k}^n \Pr(X^i>Y^{j+1-i}). \label{secondaryinequality}    
\end{equation}
If $j \leq k,$ notice that we have now already established \cref{maininequality}. Next, we consider $j > k$. Note that \[\sum_{i=1}^n \Pr(X^i > Y^{j+1-i}) = \sum_{i=1}^n \Pr(X^i \geq a_j) = \sum_{i=1}^n \frac{\Pr(X^i \geq a_j) + \Pr(Y^i \geq a_j)}{2}  = \frac{j}{2}.\]
Furthermore, notice that we can also express the first sum in \cref{secondaryinequality} as:
\begin{eqnarray*}
\sum_{i=1}^{j-k} \Pr(X^i>Y^k>X^{i+k}) &=&
\sum_{i=1}^{j-k}\sum_{t = 1}^k \Pr(X^{i+t-1}>Y^{k}>X^{i+t}) \\
&=& \sum_{t = 1}^k\sum_{i=1}^{j-k} \Pr(X^{i+t-1}>Y^{k}>X^{i+t}) \\
&=& \sum_{t = 1}^k \Pr(X^{t}>Y^{k}>X^{t +j - k}). \\
\end{eqnarray*}
Let us now introduce the notation $m = \min \{j - k, k\}$. We have that the above sum is in fact: 
\begin{eqnarray*}
 \sum_{i=1}^{m} \Pr(X^i>Y^{k}>X^{i+j-m}) &=&  \sum_{i=1}^{m} \Pr(X^i>Y^{k}) - \sum_{i=1}^m\Pr(X^{i+j-m}> Y^k) \\
 & = & \sum_{i=1}^{m} \Pr(X^i>Y^{k}) - \sum_{i=1}^m\Pr(X^{j+1-i}> Y^k)\\
& = & \sum_{i=1}^{m} \Pr(X^i>Y^{k} > X^{j+1-i})\\
&\geq & \sum_{i=1}^{m} \Pr(X^i>Y^{j+1-i}>X^{j+1-i})\\
&= & \sum_{i=1}^{m} \Pr(X^i>Y^{j+1-i}) -\Pr(X^{j+1-i} >Y^{j+1-i}).
\end{eqnarray*}
The inequality is a consequence of \cref{lemma:inbetweenrank}. We conclude that \cref{secondaryinequality} is at least:
\begin{eqnarray*}
&&\sum_{i=1}^{m} \Pr(X^i>Y^{j+1-i}) - \sum_{i=1}^{m} \Pr(X^{j+1-i}>Y^{j+1-i})+\frac{j}{4} - \frac{1}{2}\sum_{i=1}^{j-k}\Pr(X^i>Y^{j+1-i}) \\
&=& \sum_{i=1}^{m} \Pr(X^i>Y^{j+1-i}) - \frac{m}{2}+\frac{j}{4} - \frac{1}{2}\sum_{i=1}^{j-k}\Pr(X^i>Y^{j+1-i})\\
&=& \frac{1}{2}\sum_{i=1}^{k} \Pr(X^i>Y^{j+1-i}) +\frac{j-2m}{4} - \frac{1}{2}\sum_{i=m+1}^{j-m}\Pr(X^i>Y^{j+1-i})\\
&=& \frac{1}{2}\sum_{i=1}^{k} \Pr(X^i>Y^{j+1-i}) +\frac{j-2m}{4} - \frac{1}{2}\left(\frac{j-m - (m+1) + 1}{2}\right)\\
&=& \frac{1}{2}\sum_{i=1}^{k}\Pr(X^i>Y^{j+1-i}), \\
\end{eqnarray*}
which concludes the proof. Note that the penultimate line follows from the fact that:
$\Pr(X^i > Y^{j+1-i}) + \Pr(X^{j+1-i}> Y^i) = 1$ due to the symmetry between the $Xs$ and $Y$s, so the first and last terms in the sum can be paired up.
\end{proof}

\section{The $(k- \sqrt{2k\log k})^{\text{th}}$ sample as a threshold}
In this section, we will establish \cref{k-select-large}. 

\subsection{Preliminaries}

We start by stating the following useful result of \cite{CDL23}:

\begin{lemma}[\cite{CDL23}]
The algorithm for the $k$-unit prophet inequality problem which sets a deterministic price of $p$ obtains a competitive ratio of:
\[\min\left\{\Pr\left(X^k < p\right), 
  \frac{\sum_{i=1}^k \Pr(X^i \ge p)}{k}\right\}.\]
  \label{lemma:CDL}
\end{lemma}

We will also need the following lemmas. The first is due to \cite{NV23} (see Remark 1 and Lemma 2.2 in this work) and provides us with a useful way to reduce the problem to one involving just binomial random variables:
\begin{lemma}[\cite{NV23}]
If $a\leq b$, then $\Pr(X^a<Y^b)$ or equivalently $\Pr(Y^a<X^b)$ is at most $\Pr(Z<a)$, where $Z$ is a binomial random variable with $n=a+b-1$ and $p=\frac{1}{2}$.\footnote{In other words,
$\Pr(Z = a-1) = \frac{\binom{a+b-1}{a-1}}{2^{a+b-1}}.$} This is equivalent to what the probability that $X^a<Y^b$ would be if the $X_i$ were i.i.d., and $n \to \infty$. 

We will henceforth refer to this as ``the binomial case", and when we refer to $\Pr(X^a<Y^b)$ in the binomial case, we will in fact mean $\Pr(Z<a)$.
\label{Pranav Lemma}
\end{lemma}

The other lemma concerns the probability of moderate deviations in the binomial distribution with $p=\frac{1}{2}$. Its proof, which depends on a simplified version \cite{moderatedeviations} of a classical result of Cramér \cite{Cramerclassic} (see  \cite{cramér2022new} for a translation), may be found in the appendix:
\begin{lemma}
If $Z\sim \text{Binom}(r+k-1, \frac{1}{2})$, where $r=k-\sqrt{ck\log k}$, then 
$$ \Pr(Z\leq k-1) = 1- \Theta\left(\frac{k^{-\frac{c}{4}}}{\sqrt{\log k}}\right).$$
\label{lemma:binomialbound}
\end{lemma}

We are ready to prove \cref{k-select-large}.

\subsection{Proof of \cref{k-select-large}}

\begin{proof}
Note that if $x$ and $y$ are at most $1$, then $\min(x, y) \geq x+ y-1$. Therefore, by setting a price equal to $p$, \cref{lemma:CDL} guarantees a competitive ratio of at least:
\[\Pr\left(X^k < p\right)+ 
  \frac{\sum_{i=1}^k \Pr(X^i \ge p)}{k}-1,\]
and therefore by letting $p$ be random and equal $Y^r$ for some $r \leq k$, we obtain a competitive ratio of at least:
\begin{align*}
\Pr\left(X^k < Y^r\right)+ 
  \frac{\sum_{i=1}^k \Pr(X^i > Y^r)}{k}-1 &=
  \frac{\sum_{i=1}^k \Pr(X^i > Y^r > X^k)}{k}\\
  &\geq \frac{\sum_{i=1}^r \Pr(X^i > Y^r > X^k)}{k}\\
  &= \frac{-r\Pr(X^k > Y^{r}) + \sum_{i=1}^{r}\Pr(X^i>Y^{r})}{k}.
\end{align*}
We see that we may safely apply Lemma \ref{Pranav Lemma} to all the probabilities in this expression, so we will now work with probabilities in the binomial case. Notice that by summation-by-parts, we must have:
\[\sum_{i=1}^{r}\Pr(X^i>Y^{r} > X^k) = \sum_{j = 1}^{r -1}j\Pr(X^j > Y^{r} > X^{j+1}) + r\Pr(X^{r} > Y^{r}>X^k).\]
First note that for the second quantity on the right, we have:
\begin{align*}
r\Pr(X^{r} > Y^{r}>X^k) &= r(1-\Pr(X^{r}<Y^{r})-\Pr(Y^{r}<X^k)) \\&= r\left(\Pr(X^k < Y^{r}) - \frac{1}{2}\right)\\
&= r\left(\Pr(Z \leq k-1) -\frac{1}{2}\right)
\end{align*}  where $Z\sim \text{Binom}(r+k-1, \frac{1}{2}).$ 
Next, let us calculate the sum:
\begin{align*}
    \sum_{j=1}^{r-1}j\Pr(X^j>Y^{r}>X^{j+1}) =\sum_{j=1}^{{r}-1}\frac{j\binom{j+r-1}{j}}{2^{j+r}}
    &=r\sum_{j=1}^{r-1}\frac{\binom{j+r-1}{j-1}}{2^{j+r}}\\
    &=r\sum_{j=0}^{r-2}\frac{\binom{j+r}{j}}{2^{j+r+1}}\\
    &=r\left(\sum_{j=0}^{r}\frac{\binom{j+r}{j}}{2^{j+r+1}}-\frac{\binom{2r-1}{r-1}}{2^{2r}}-\frac{\binom{2r}{r}}{2^{2r+1}}\right)\\
    &=r\left(\sum_{j=0}^{r}\Pr(X^j > Y^{r+1} > X^{j+1})-\frac{\binom{2r-1}{r}}{2^{2r-1}}\right)\\
    &=r\left(\Pr(Y^{r+1}>X^{r+1})-\frac{\binom{2r-1}{r}}{2^{2r-1}}\right)\\
    &=r\left(\frac{1}{2}- \frac{\binom{2r-1}{r}}{2^{2r-1}}\right)
\end{align*}

Thus, we obtain a competitive ratio of at least:
\[\frac{r\Pr(Z \leq k-1) - r\frac{\binom{2r-1}{r}}{2^{2r-1}}}{k},\]
where $Z\sim \text{Binom}(r+k-1, \frac{1}{2}),$ as desired. Now,
a standard bound on binomial coefficients provable using Stirling's approximation tells us that:
\[r\frac{\binom{2r-1}{r}}{2^{2r-1}} =O(\sqrt r).\]
Furthermore, if $r = k -\sqrt{2k\log k}$, \cref{lemma:binomialbound} implies that:
\[\Pr(Z \leq k-1) = 1 - \Theta\left(\frac{1}{\sqrt{k\log k}}\right),\] and so a simple calculation gives the desired competitive ratio.
\end{proof}

\section{Impossibility of beating $1 - \sqrt{\frac{2\log k}{k}}$}
\label{sec:upperbounds}
To establish \cref{tightness}, informally speaking, we need to explain both why it might be bad to set too \textit{low} a price, and why it might be bad to set too \textit{high} a price. 

On the one hand, setting too high a price is potentially dangerous because it is possible that all random variables are roughly equally valuable. The only thing that then matters is how many items are sold. Setting too high a price means that there is the possibility of not selling enough. On the other hand, setting too low a price is potentially dangerous because it is possible that the last random variable has some small chance of being really valuable, and by setting too low a price, we sell too much and destroy our ability to exploit the last arriving random variable. We formalize these ideas in our proof of \cref{tightness}:

\begin{proof}[Proof of \cref{tightness}]

Suppose an algorithm observes the samples $\{y_1, y_2, \ldots y_n\}$, and sets a potentially random but static price $R(y_1, y_2, \cdots, y_n)$. For convenience, we will assume throughout that $n \gg k$. Suppose this algorithm, by way of contradiction, achieves a competitive ratio strictly better than $1-\sqrt{\frac{(2-\epsilon)\log k}{k}}$ for some $\epsilon>0$, and for some large $k$. 

Denote by $Z$ the random variable representing the minimum of $k$ and the number of $y$'s that exceed the threshold $R$, i.e.,
\[Z(y_1, y_2, \ldots, y_n) = \min \{k, \sum_{i=1}^n \mathds{1}({y_i \geq R})\}.\]

Suppose that all the values $y_1, y_2, \ldots, y_n$ are distinct, and suppose that these values are all in the interval $[1-\delta, 1]$, with $\delta$ small. We will see later that it will be convenient to assume:

\[\delta \leq \frac{\epsilon}{16\sqrt{k}}.\]

One possibility the algorithm must account for is that the $y_i$ are drawn from distributions that have disjoint support, and these distributions are uniform distributions centered about $y_i$ with potentially arbitrarily small radius. In this case, we are guaranteed that $\mathbb{E}\left[\sum_{i=1}^k X^i\right]\geq  k(1-\delta)$, while $\mathbb{E}[\text{ALG}] \leq \mathbb E [Z(y_1, y_2, \ldots, y_n)]$. So it must be true that:

\[\mathbb E [Z(y_1, y_2, \ldots, y_n)] \geq (k-\sqrt{(2-\epsilon)k\log k})\cdot(1-\delta).\]

Since, by definition, $Z\leq k$, we have by Markov's inequality that for any $s < k$:
\begin{eqnarray*}
  & \frac{\mathbb E [k - Z]}{k-s} &\geq \Pr(k - Z > k - s) \\
\implies &\frac{\mathbb E [Z] - s}{k-s} &\leq \Pr(Z> s).
\end{eqnarray*}
In particular, letting $s = k-\sqrt{\left(2-\frac{\epsilon}{2}\right)k\log k}$, we obtain:
\begin{eqnarray*}
    &&\Pr\left(Z(y_1, y_2, \ldots, y_n) > k-\sqrt{\left(2-\frac{\epsilon}{2}\right)k\log k} \right)\\
    &\geq &\frac{k-\sqrt{(2-\epsilon)k\log k}-\delta \cdot (k-\sqrt{(2-\epsilon)k\log k})  - k + \sqrt{\left(2-\frac{\epsilon}{2}\right)k\log k}}{\sqrt{\left(2-\frac{\epsilon}{2}\right)k\log k}} \\
    & = &\frac{\sqrt{2-\frac{\epsilon}{2}}-\sqrt{2-\epsilon}}{\sqrt{2-\frac{\epsilon}{2}}} - \frac{\delta \cdot (k-\sqrt{(2-\epsilon)k\log k})}{\sqrt{\left(2-\frac{\epsilon}{2}\right)k\log k}} \\
    & \geq &\frac{\epsilon}{8} - \delta\sqrt{k} \\
    &\geq & \frac{\epsilon}{16}.
\end{eqnarray*}
We conclude that
\begin{align*}
\Pr (Z(y_1, y_2, \ldots, y_n) > s)&\geq \frac{\epsilon}{16},
\end{align*}
for \textit{every} distinct $y_1, y_2, \cdots, y_n \in [1-\delta, 1]$.

Another set of possibilities the algorithm needs to account for is that the $y_i$ are generated from distributions in the following way: $y_i \sim \text{Unif}(1-\delta,1)$, and $y_n \sim \text{Unif}(1-\delta,1)$ with probability $\frac{N-1}{N}$ and $N^2$ with probability $\frac{1}{N}$. We have then that the competitive ratio in this setting, as $N$ tends to infinity, tends to one minus the probability that the algorithm makes $k$ picks out of $X_1, \ldots, X_{n-1}$.

Now we know from lemma \ref{Pranav Lemma} and lemma \ref{lemma:binomialbound} that in the limit as $n \to \infty$ that
$$\Pr(X^k>Y^s) = \Theta\left(\frac{k^{\frac{-(2-\epsilon/2)}{4}}}{\sqrt{\log k}}\right),$$
where $X^k$ represents the $k$th largest amongst $X_1, \ldots, X_{n-1}$, and $Y^s$ is definied similarly. Furthermore, the threshold, and thus the value of $Z$, is chosen before the $X_i$ are realized. It follows that
\[ \Pr(X^k>Y^s, Z > s)\geq \frac{\epsilon}{16}\cdot \Pr(X^k>Y^s) = \Theta\left(\frac{k^{\frac{-(2-\epsilon/2)}{4}}}{\sqrt{\log k}}\right),\]
since regardless of the value of $Y^s$, $Z$ has an $\frac{\epsilon}{16}$ chance of being bigger than $s$. In case $X^k> Y^s$ and $Z >s$, more than $k$ of the $X_1, \ldots, X_{n-1}$ surpass the threshold, and thus, as $n$ and $N$ grow very large, the competitive ratio is less than or equal to
\[1-\Theta\left(\frac{k^{\frac{-(2-\epsilon/2)}{4}}}{\sqrt{\log k}} \right),\]
which contradicts our assumption that we were working with an algorithm whose competitive ratio surpassed $1-\sqrt{\frac{(2-\epsilon)\log k}{k}}$.\end{proof}

\printbibliography

\appendix
\section{Omitted Proofs}

\subsection{Proof of Lemma \ref{1/2 Lemma}}

   Consider the quantity $\Pr(X^i>Y^{j+1-i}, Y^k>X^{i+k})=\Pr(X^i>Y^{j+1-i})\Pr(Y^k>X^{i+k}|X^i>Y^{j+1-i})$. Let us concentrate on the conditional probability. We know that there is a certain number $\theta$ of $Y$'s, $\theta \leq j-i$, greater than $X^i$. We have also that $Y^k>X^{i+k}$ if the sequence $a_1,\ldots,a_{i+2k-1}$ contains at least $k$ $Y$'s. This will hold if $a_{i+\theta + 1},\ldots,a_{i+2k-1}$ contains at least $k-\theta$ $Y$'s. Consider revealing the $\theta$ $X$'s sampled from the same distribution as $Y^1,\ldots,Y^\theta$. At most $\theta$ spots amongst $a_{i+\theta+ 1},\ldots,a_{i+2k-1}$ have been taken up by these $X$'s. The remaining spots, of which there are $2k - 2\theta-1$, each have at least a $\frac{1}{2}$ chance of being a $Y$ value, and we would like  $k-\theta$ of these spots to be $Y$'s. If each spot had exactly a $\frac{1}{2}$ chance of being a $Y$, then the chance of $k - \theta$ $Y$'s would be exactly $\frac{1}{2}$, so we conclude $\Pr(Y^k>X^{i+k}|X^i>Y^{j+1-i})\geq \frac{1}{2}.$

\subsection{Proof of Lemma \ref{lemma:inbetweenrank}}
In preparation for our proof of the lemma, we have the following claim:
\begin{claim}\label{claim}
If $i< j < k$, then \[\Pr(X^{j-1}> Y^i> X^{j})\geq \Pr(X^{j}> Y^i> X^{j+1}) \text{ and }\Pr(X^{j-1}> Y^k> X^{j})\leq \Pr(X^{j}> Y^k> X^{j+1}).\]
\end{claim}

\begin{proof}
Let us concentrate on the first inequality. The proof of the second inequality is completely analogous. Let us say that $a_r$ and $a_s$ are a pair if they come from the same distribution. Let us focus on the pairs among $a_1, a_2, \ldots, a_{i+j-1}$. Say there are $l$ of them. This means that of $a_1\dots a_{i+j-2}$, either $i+j-2l-1$ (if $a_{i+j-1}$ is paired) or $i+j-2l-2$ otherwise, are unpaired and therefore assigned uniformly at random to $X$’s or $Y$’s. The event $X^{j-1}>Y^i>X^j$ requires exactly $j-1-l$ of those being $X$’s. If this occurs, there is a $\frac{1}{2}$ chance that $a_{i+j-1}$ is a $Y$, regardless of whether it is paired. Thus: 
\[\Pr(X^{j-1}>Y^i> X^{j}) = \begin{cases}
  \frac{1}{2^{i+j-2l-1}}\binom{i+j-2l-2}{j-l-1}  & \text{ if } a_{i+j-1} \text{ is unpaired}\\\
  \frac{1}{2^{i+j-2l}}\binom{i+j-2l-1}{j-l-1}  & \text{ if } a_{i+j-1} \text{ is paired}
\end{cases}\]
This is Claim 1 in \cite{NV23}. Note that:
\begin{alignat*}{5}
&\frac{1}{2^{i+j-2l}}&\binom{i+j-2l-1}{j-l-1} \quad &\geq&&\quad \frac{1}{2^{i+j-2l-1}}\binom{i+j-2l-2}{j-l-1}\\
\iff &  &\binom{i+j-2l-1}{j-l-1}\quad  &\geq&&\quad  2\binom{i+j-2l-2}{j-l-1}\\
\iff &  &i+j-2l-1 \quad &\geq&&\quad  2(i-l)\\
\iff &  &j-1 \quad &\geq&&\quad i
\end{alignat*}
We conclude that $\Pr(X^{j-1}>Y^i> X^{j})$ is smaller when $a_{i+j-1}$ is unpaired, and $\Pr(X^{j}>Y^i> X^{j+1})$ is smaller when $a_{i+j}$ is unpaired. So to establish that $\Pr(X^{j-1}> Y^i> X^{j})\geq \Pr(X^{j}> Y^i> X^{j+1})$, we may concentrate on the case when $a_{i+j-1}$ is unpaired (with $l$ pairs among $a_1, a_2, \ldots, a_{i+j-1}$), and $a_{i+j}$ is paired (with $l+1$ pairs among $a_1, a_2, \ldots, a_{i+j}$), but in this case, in fact, the inequality is tight:
\[\frac{1}{2^{i+j-2l-1}}\binom{i+j-2l-2}{j-l-1} = \frac{1}{2^{i+(j+1)-2(l+1)}}\binom{i+(j+1)-2(l+1)-1}{j+1-(l+1)-1}.\]
This completes the proof.
\end{proof}

We are ready to prove \cref{lemma:inbetweenrank}:
\begin{proof}
For the sake of convenience, let us introduce the notation $F(i, j, k) := \Pr(X^i > Y^j > X^k)$, so the goal is to show $F(i, j, k) \geq F(i, k, k)$ in case $i \leq j \leq k$. Note that if $j = k$, there is nothing to prove. Furthermore, observe:
\[    F(i, k, k) = \Pr(X^i> Y^k) - \Pr(X^k>Y^k) = \Pr(Y^i> X^k) - \Pr(Y^i>X^i) = F(i, i, k).\]
In particular, there is nothing to prove if $j = i$, so let us assume $i < j < k$. Note now that \cref{claim} implies:
\[F(j-1, i, j)F(j, k, j+1) \geq F(j-1, k, j)F(j, i, j+1).\]
Hence, we conclude that in case $F(j-1, k, j) > F(j-1, i, j)$, it must be that $F(j, k, j+1) > F(j, i, j+1)$. But:
\begin{alignat*}{3}
&F(j-1, k, j) - F(j-1, i, j)\quad= \quad F(i, j-1, k) - F(i, j, k) \\
\text{and }& F(j, k, j+1) - F(j, i, j+1)\quad= \quad F(i, j, k) - F(i, j+1, k)
\end{alignat*}
so we conclude that if $i < j < k$, then in case $F(i, j-1, k) > F(i, j, k)$, it must follow that $F(i, j, k) > F(i, j+1, k)$. Suppose now by way of contradiction that:
\[j^* = \min{\text{argmin}_{j \in [i, k]}F(i, j, k)} \neq i.\]
Then, $F(i, j^*-1, k) > F(i, j^*, k)$, and $F(i, j^*, k) \leq F(i, j^*+1, k),$ which is a contradiction.
\end{proof}

\subsection{Proof of Lemma \ref{lemma:binomialbound}}

The following proposition from \cite{moderatedeviations} is a simplification of a classical result of Cramér \cite{Cramerclassic}:

\begin{proposition}[\cite{moderatedeviations}]
    If $X_1, \ldots, X_n$ are independent random variables, $1$ or $-1$ uniformly at random, $n\lambda_n^3 \to 0$, and $n\lambda_n^2 \to \infty$, then 
    \begin{equation*}
        \Pr\left(\left|\frac{\sum X_i}{n}\right|>\lambda_n \right)\sim \frac{2}{\sqrt{2\pi n\lambda_n^2}}e^{\frac{-n\lambda_n^2}{2}}.
    \end{equation*}
    \label{proposition:Cramer}
\end{proposition}
We use the proposition to establish Lemma \ref{lemma:binomialbound}:

\begin{proof}
    First, we re-define $X_1, \ldots, X_n$ to be $0$ or $1$, uniformly at random, meaning that the left side of Proposition \ref{proposition:Cramer} becomes 
    $$\Pr\left(\left|\frac{\sum X_i}{n}-\frac{1}{2}\right|>\frac{\lambda_n}{2} \right).$$
    This is equal, by algebra and symmetry, to 
    $$\frac{1}{2}\Pr\left(\sum_{i=1}^n X_i-\frac{n}{2}>\frac{n\lambda_n}{2} \right),$$
    which, if $Z\sim \text{Binom}(n, \frac{1}{2})$, is 
    $$ \frac{1}{2}\Pr\left(Z>\frac{n}{2}+\frac{n\lambda_n}{2}\right). $$
    Therefore, if we let $r=k-\sqrt{ck\log k}$, $n=r+k-1$, and $n\lambda_n = \sqrt{ck\log k}-1$, take $k$ to infinity, and apply the proposition, we get 
    \begin{align*}
        \frac{1}{2}\Pr\left( Z>k-1\right) &= \frac{1}{2}\Pr\left( Z>\frac{r+k-1}{2} +\frac{1}{2}\sqrt{ck\log k}-\frac{1}{2}\right) \\
        &= \Theta\left(\frac{\sqrt{n}}{n\lambda_n}\exp\left({\frac{-n^2\lambda_n^2}{2n}}\right)\right) \\
        &= \Theta\left(\frac{1}{\sqrt{\log k}}\exp\left({\frac{-n^2\lambda_n^2}{2n}}\right)\right) \\
        &= \Theta\left(\frac{1}{\sqrt{\log k}}\exp\left({\frac{-(\sqrt{ck\log k}-1)^2}{2(r + k -1)}}\right)\right). \\
    \end{align*}
    But notice that,
    \begin{align*}
       \left|\frac{-(\sqrt{ck\log k}-1)^2}{2(r + k -1)} - \frac{-c\log k}{4}\right| &= \left|\frac{(r+k-1)c\log k-2(\sqrt{ck\log k}-1)^2}{4(r + k -1)}\right| \\
        &\leq \left|\frac{(r+k-1)c\log k-2ck\log k - 2 + 4\sqrt{ck\log k}}{4k}\right|\\
        &= \left|\frac{(r-k-1)c\log k - 2 + 4\sqrt{ck\log k}}{4k}\right|\\
        &= \left|\frac{(\sqrt{ck\log k}+1)c\log k + 2 - 4\sqrt{ck\log k}}{4k}\right|,
    \end{align*}
    and clearly the last expression is less than a constant.
    We conclude that:
    \begin{align*}
        \Pr\left( Z>k-1\right) &= \Theta\left(\frac{1}{\sqrt{\log k}}\exp\left(\frac{-c\log k}{4}\right)\right). \\
    \end{align*}
    
    This implies the result: 
    \begin{equation*}
        \Pr\left(Z\leq k-1\right)=1-\Theta\left(\frac{k^{\frac{-c}{4}}}{\sqrt{\log k}}\right).
    \end{equation*}
\end{proof}

\subsection{Proof that \cite{CDL23} obtains $1-\sqrt{\frac{\log k }{k}}-o\left(\sqrt{\frac{\log k}{k}}\right)$}
\label{sec:old-result}
In the case where the distributions are known, \cite{CDL23} gives a static threshold algorithm that can achieve a competitive ratio given by either side of the equation 
$$ \frac{\EX[\min(X,k)]}{k}=\Pr(X\leq k-1),$$
where $X$ is a Poisson random variable with mean $\lambda_k$ set so that both sides of the above equation are equal. \cite{JMZtight} showed that this algorithm is in fact optimal, that it has a competitive ratio that is $1 - \Theta\left(\sqrt{\frac{\log k }{k}}\right)$, and $\lambda_k = k - \Theta\left(\sqrt{k\log k} \right)$. Here, we will show that this competitive ratio is in fact
$$ 1-\sqrt{\frac{\log k }{k}}-o\left(\sqrt{\frac{\log k}{k}}\right).$$
(Most of the proof of this result is already present in \cite{JMZtight}, but we include a slightly different proof for the sake of completeness.) Let $Z_c$ be a Poisson random variable with mean $k - \sqrt{ck\log k}$ for a constant $c$.

First of all, note that we can provide a good estimate for $\frac{\EX[\min(Z_c,k)]}{k}$. Indeed, 
\[1-\sqrt{\frac{c\log k}{k}} = \frac{\EX[Z_c]}{k} \geq \frac{\EX[\min(Z_c,k)]}{k},\]
and furthermore,
\begin{align*}
 \frac{\EX[\min(Z_c,k)]}{k}=\frac{\EX[Z_c-(Z_c-k)1_{Z_c\geq k}]}{k}&\geq \frac{\EX[Z_c]}{k}-\frac{\EX[(Z_c-\EX[Z_c])1_{Z_c \geq k}]}{k}\\&\geq \frac{\EX[Z_c]}{k}-\frac{\EX[|Z_c-E[Z_c]|]}{k}\\
    &\geq \frac{\EX[Z_c]}{k}-\frac{\sqrt{\EX[(Z_c-\EX[Z_c])^2]}}{k}\\
    &= \frac{\EX[Z_c]}{k}-\frac{\sqrt{\EX[Z_c]}}{k} \\
    &\geq 1-\sqrt{\frac{c\log k}{k}}-\frac{1}{\sqrt{k}},
\end{align*}
where $\EX[(Z_c-\EX[Z_c])^2] = \EX[Z_c]$ since $Z_c$ is Poisson. Second of all, \cite{JMZtight} (in their proof of Proposition 3) show that for a constant $c$:
\[\Pr(Z_c \geq k) = \Theta\left(\frac{\exp(-c \log k/2)}{\sqrt{c\log k}}\right) = \Theta\left(\frac{k^{\frac{-c}{2}}}{\sqrt{\log k}}\right).\]
In fact, the statement is proved in \cite{JMZtight} as a direct consequence of Corollary 7.2 in \cite{Harremoës2016} and the fact that $
1 - \Phi(x) = \Theta\left(\frac{\exp(-x^2/2)}{x \sqrt{2\pi}}\right)
$, where $\Phi$ is the CDF of the normal distribution. Now if $\lambda_k > k - \sqrt{ck \log k}$ for some constant $c < 1$ for infinitely many $k$, then,
\[\frac{\EX[\min(X,k)]}{k}>  \frac{\EX[\min(Z_c,k)]}{k}\geq 1-\sqrt{\frac{c\log k}{k}}-\frac{1}{\sqrt{k}} \overset{\text{large }k}{\geq} \Pr(Z_c \leq k-1) > \Pr(X \leq k-1),\]
which is a contradiction. On the other hand, if  $\lambda_k < k - \sqrt{ck \log k}$ for some constant $c > 1$ for infinitely many $k$, then
\[\frac{\EX[\min(X,k)]}{k}< \frac{\EX[\min(Z_c,k)]}{k}\leq 1-\sqrt{\frac{c\log k}{k}}\overset{\text{large }k}{\leq} \Pr(Z_c \leq k-1) < \Pr(X \leq k-1),\]
which is also a contradiction. The desired result follows.

\end{document}